\documentclass[11pt,onecolumn,amssymb,nofootinbib]{revtex4}
\usepackage{bm}
\usepackage{amsmath}
\usepackage{amssymb}
\usepackage{graphicx}
\usepackage{amsmath}
\usepackage{bbm}

\begin{document}
\title{\bf The Conway-Kochen Argument and relativistic GRW Models}
\author{Angelo Bassi}
\email{bassi@mathematik.uni-muenchen.de} \affiliation{Mathematisches
Institut der Universit\"at M\"unchen, Theresienstr. 39, 80333
M\"unchen, Germany, \\The Abdus Salam International Centre for
Theoretical Physics, Strada Costiera 11, 34014 Trieste, Italy.}
\author{GianCarlo Ghirardi}
\email{ghirardi@ts.infn.it} \affiliation{Department of Theoretical
Physics and I.N.F.N., Strada Costiera 11, 34014 Trieste, Italy,
\\The Abdus Salam International Centre for Theoretical Physics,
Strada Costiera 11, 34014 Trieste, Italy.}
\begin{abstract}
$ $  \\ \\ In a recent paper, Conway and Kochen proposed what is now
known as the ``Free Will theorem'' which, among other things, should
prove the impossibility of combining GRW models with special
relativity, i.e., of formulating relativistically invariant models
of spontaneous wavefunction collapse. Since their argument basically
amounts to a non-locality proof for any theory aiming at reproducing
quantum correlations, and since it was clear since very a long time
that any relativistic collapse model must be non-local in some way,
we discuss why the theorem of Conway and Kochen does not affect the
program of formulating relativistic GRW models.
\end{abstract}
\maketitle

\section*{1. THE CONWAY-KOCHEN ARGUMENT}

We briefly review the argument by Conway and Kochen \cite{ck}.
Following their way of presenting it, we do not assume any
particular theory underlying physical phenomena; we only assume that
certain particular physical systems exist---independently of how
they are described by any theory---upon which measurements can be
made; more specifically, we assume that there exist systems which we
call ``particles of spin 1'', upon which operations which we call
``measurement of the square of the spin along the direction $n$''
can be performed ($n$ denotes any direction in the three-dimensional
space); we call $S^{2}_{n}$ the outcome of such a kind of
measurement along the direction $n$. We assume that when $x,y,z$
represent an orthogonal triple of directions, the three
corresponding measurement-operations can be simultaneously performed
on the system\footnote{This property reflects the well known fact
that, in the case of a spin 1 particle, the squares of the  spin
operators along three orthogonal directions commute among
themselves.}; we assume furthermore, in agreement with Quantum
Mechanical rules, that:
\begin{equation} \label{eq:prop-spin}
\begin{array}{l}
\makebox{$S^{2}_{n}$ takes only the value $0$ or $1$, for any $n$;}
\\
\makebox{For any orthogonal triple $x,y,z$, one has: $S^{2}_{x} +
S^{2}_{y} + S^{2}_{z} = 2$,}
\end{array}
\end{equation}
whenever the three corresponding measurements are (simultaneously)
performed. Trivially, properties~\eqref{eq:prop-spin} imply that one
of $S^{2}_{x}$, $S^{2}_{y}$, $S^{2}_{z}$ is 0, while two are 1.

We finally assume the standard formalism of special relativity, in
particular concepts like the backward light-cone and space-like
separated regions of spacetime.

\subsection*{1.1. The Postulates}

Conway and Kochen consider the following three axioms:

\noindent {\bf TWIN:} It is possible to produce two spin 1 particles
which are in the state of ``total spin 0'', meaning with this that
if a measurement of the square of the spin along the direction $n$
is performed on one particle, giving the outcome $S^{2}_{n}$, then a
measurement of the square of the spin along the same direction $n$
performed on the other particle gives the same outcome $S^{2}_{n}$.
Moreover, such a property does not depend on the relative position
of the two particles and on the relative time on which the two
experiments are performed; in particular, it holds true when the two
measurements are space-like separated.

\noindent {\bf FREE:} Each experimenter can freely choose any
direction $n$ along which to perform the measurement.

\noindent {\bf FIN:} Information cannot travel at a speed greater than
the speed of light.

In the last axiom, we used the term ``information'' in an intuitive
sense, without specifying what it means; though we do not like to
resort to such a vague term in setting the axioms of any logical
reasoning, we use it simply to adhere to the original formulation of
\cite{ck}.

\subsection*{1.2. The Argument}

Let us consider two spin 1 particles which are in a state of total
spin 0; let us label with ``$a$'' one of the two particles and with
``$b$'' the other one, and with ``$A$'' an experimenter who performs
a measurement on $a$, and with ``$B$'' one who performs a
measurement on $b$. Let $n$, $m$ and $\ell$ be three orthogonal axis
along which $A$ decides to perform three practically simultaneous
measurements of the square of the spin of particle $a$.

We assume that the outcome of the measurement of the square of the
spin of particle $a$ along the direction $n$ is a function of all
the information $\alpha$ contained in the backward light-cone of the
particle (with respect to the spacetime point where the measurement
is performed), and of the other two directions\footnote{The
necessity  of allowing, in principle, that the outcome along $n$
depends on the other chosen directions follows from the so-called
Kochen-Specker  theorem \cite{ks}.} $m$ and $\ell$ chosen by $A$:
\begin{equation}
S^{2}_{a:n} \quad \equiv \quad S^{2}_{a:n}(m,\ell;\alpha)
\end{equation}
(we have slightly changed the notation from $S^{2}_{n}$ to
$S^{2}_{a:n}$ in order to distinguish when $S^{2}_{n}$ refers to a
measurement performed on particle $a$ and when it refers to a
measurement performed on particle $b$, along the direction $n$).

In a similar way, if $B$ decides to simultaneously measure the
square of the spin of particle $b$ along three orthogonal directions
$i,j,k$, we assume that the outcome $S^{2}_{i}$ may depend only on
the information $\beta$ contained in the backward light-cone and on
the other two directions $j,k$ chosen by $B$:
\begin{equation}
S^{2}_{b:i} \quad \equiv \quad S^{2}_{b:i}(j,k;\beta)
\end{equation}
Assume finally that the two experimenters perform their measurements
at space-like separated regions, so that there cannot be any
exchange of information among them before the two measurements are
over. The argument now goes as follows.

Because of {\bf TWIN}, if $A$ and $B$ choose a common axis $n$, then
the two outcomes must be perfectly correlated:
\begin{equation}
S^{2}_{a:n}(m,\ell;\alpha) \quad = \quad S^{2}_{b:n}(j,k;\beta),
\end{equation}
for any direction $n$. But, because of {\bf FREE}, $B$ can choose to
perform his measurement along any orthogonal triple $n,k,j$, with
$n$ fixed. Because of {\bf FIN}, such a choice cannot affect
$S^{2}_{a:n}(m,\ell;\alpha)$. This means that the outcome of the
spin  measurement along $n$ cannot depend on which other two
orthogonal directions $j,k$ are chosen by  the experimenter, i.e. :
\begin{equation}
S^{2}_{b:n}(j,k;\beta) \quad = \quad S^{2}_{b:n}(\beta).
\end{equation}
Obviously, a similar conclusion holds for
$S^{2}_{a:n}$.

The contradiction now arises because, as it is well known, a
function like $S^{2}_{a:n}(\alpha)$ or $S^{2}_{b:n}(\beta)$ cannot
exists \cite{ks}. As a consequence, the outcome $S^{2}_{a:n}$ of the
measurement of the square of the spin of particle $a$ (or $b$)
cannot be uniquely determined by the past information contained in
the backward light-cone from the region where A (or B) performs  its
measurement.

\subsection*{1.3. Comments}

The provocative conclusion of Conway and Kochen is that the
particles' response to the experiment is free, i.e. that the outcome
of an experiment cannot be entirely determined by the previous
information accessible to whom performs the measurement, but indeed
the moral stemming out of the above argument is well known and less
surprising: {\it no local theory exists which fully agrees with
quantum mechanical predictions.} As a matter of fact, what Conway
and Kochen have shown is that the three postulates listed above,
plus the existence of a functional relation between the outcomes of
certain spin experiments and certain ``information'', lead to a
contradiction, from which they conclude that such a functional
relation cannot exist. But indeed, after the work of Bell it is well
known that the conclusion is a different one: Nature is non-local,
i.e. {\bf FIN} is wrong, if, as the authors seem to suggest,
information---apart from its ambiguous meaning---includes everything
which might possibly determine an event (in our case, the outcome of
a certain experimental procedure). And indeed, the above argument
was first proposed by P. Heywood and M.L.G. Redhead \cite{hr} and by
A. Stairs \cite{sta}, further explored by H.R. Brown and G.
Svetlichny \cite{bro} and subsequently generalized by A. Elby
\cite{elb} (see also \cite{hem}); in these papers the above theorem
is correctly presented as a non-locality proof, its novelty being
that it differs from the original proof by Bell, by combining ideas
previously related only to contextuality\footnote{We thank S.L.
Adler and D. D\"urr for having brought the above papers to our
attention.}.

The reason why after Bell's work one has to conclude that Nature is
non-local is the following: Bell's theorem does not require the
existence of a functional relation between the outcomes of
experiments and past information. It simply requires the {\bf FREE}
assumption, the analog of the {\bf TWIN} axiom for $1/2$-spin
particles, and Bell's definition of locality, which in some sense is
the analog of the {\bf FIN} axiom, even though it is expressed in
clearer mathematical and physical terms. From these three
axioms---we insist: without assuming any other functional
dependence---one can derive an inequality which turns out to be
violated by Nature. Accordingly, one of the three axioms must be
wrong. Since {\bf TWIN} has been experimentally verified and no-one
is willing to deny {\bf FREE}, then Bell's locality must be
violated: Nature, in a very precise sense, is non-local.

Here we do not want to  discuss  the merits of the different proofs,
whether Bell's definition of locality is more or less general than
those which have been subsequently proposed, included the above {\bf
FIN} axiom; the moral is basically the same as the one given by
Bell's theorem: any theory, in one way or another, must be non-local
if it has to be empirically equivalent to Quantum Mechanics.

We stress it once more: the series of experiments performed by A.
Aspect \cite{As1,As2} proved that Nature, by violating Bell's
inequalities, violates Bell's condition of locality: this means, as
far as we understand physics now, that Nature is non-local, that
something happening in some region of space can affect the state of
far away systems. On the other hand, as proven in \cite{Eb,G} for a
quantum theory with the reduction postulate, such non-locality
cannot be used to signal at a speed greater than the speed of light;
this is what has been called the ``peaceful coexistence'' between
Quantum Mechanics and Special Relativity \cite{Sh}, which renders
Quantum Mechanics compatible with Special Relativity in spite of its
non-local character.

People working on collapse models were of course aware of this
non-locality constraint which any collapse model, whether
relativistic or not, has to obey in order to be compatible with
Quantum Mechanics; thus, for them, the argument of Conway and Kochen
does not come as a surprise.

But there is something more. In applying their theorem to GRW
models, Conway and Kochen mistakenly assume that the response of a
particle (i.e. the outcome of a measurement) may depend {\it only}
on the jumps which occurred in the backward light-cone of the
spacetime point where the measurement occurs, since they regard the
jumps as information which must fulfill {\bf FIN}. But this cannot
possibly be the right picture, even at a relativistic level, if the
GRW model is to account for the nonlocal features of entangled
quantum systems, which have been elucidated by Bell's work; we now
discuss this issue in more detail.

\section*{2. MODELS OF SPONTANEOUS WAVEFUNCTION COLLAPSE}

We briefly review some of the main features of the GRW model of
spontaneous wavefunction collapse, which are relevant for the
present discussion. We will first present the non relativistic GRW
model \cite{grw} and then comment on possible relativistic
generalizations.

\subsection*{2.1. The GRW Model}

The starting point of the GRW model is that the wavefunction alone
is the complete mathematical description of all physical phenomena,
it represents the maximum knowledge one can have, in principle,
about the state of a physical system, both microscopic and
macroscopic. Since macroscopic objects are always well localized in
space, while the Schr\"odinger equation allows for superpositions of
different macroscopic states, one has to modify the standard quantum
evolution in order to provide a consistent and unified description
of micro- and macro-phenomena. This is done in the following way.

Let us consider a system of $N$ particles; let ${\mathcal H}$ be the
Hilbert space associated to it and $H$ the standard quantum
Hamiltonian of the system. The model is defined by the following postulates:

\noindent 1. At  random times, each particle experiences a sudden
jump of the form:
\begin{equation}
\psi_{t} \quad \longrightarrow \quad \frac{L_{n}({\bf x})
\psi_{t}}{\|L_{n}({\bf x}) \psi_{t}\|},
\end{equation}
where $\psi_{t}$ is the statevector of the whole system at time $t$,
immediately prior to the jump process. $L_{n}({\bf x})$ is a linear
operator which is conventionally chosen equal to:
\begin{equation}
L_{n}({\bf x}) \quad = \quad \sqrt[4]{\left(
\frac{\alpha}{\pi}\right)^3} \exp \left[ - \frac{\alpha}{2} ({\bf
q}_{n} - {\bf x})^2 \right],
\end{equation}
where $\alpha$ is a new parameter of the model which sets the the
width of the localization process, and ${\bf q}_{n}$ is the
position operator associated to the $n$-th particle. The random
variable ${\bf x}$ corresponds to the place where the jump occurs.

\noindent 2. Between two consecutive jumps, the statevector evolves
according to the standard Schr\"odinger equation.

\noindent 3. The probability density for a jump taking place at
the position ${\bf x}$ is given by:
\begin{equation}
p_{n}({\bf x}) \quad \equiv \quad \|L_{n}({\bf x}) \psi_{t}\|^2;
\end{equation}
the probability density for the different particles are independent.

\noindent 4. Finally, it is assumed that the jumps are distributed
in time according to a Poissonian process with frequency
$\lambda$, which is the second new parameter of the model.

The standard numerical values for $\alpha$ and $\lambda$
are\footnote{Recently S.L. Adler proposed a radically different
numerical value for the collapse rate $\lambda$; see ref.
\cite{adl}.}:
\begin{equation}
\lambda \; \sim \; 10^{-16} \, \makebox{sec$^{-1}$} \quad\qquad
\alpha \; \sim \; 10^{-10} \, \makebox{cm$^{-2}$},
\end{equation}
which have been chosen in such a way to guarantee a very good
agreement of GRW with standard quantum mechanics and, at the same
time, to ensure an almost instantaneous localization of the
wavefunction of classical macro-objects, thus suppressing the
unwanted superpositions of differently located macro-states.

The evolution being stochastic, any initial state $\psi_{0}$ evolves
in time into an ensemble of states $\{ \psi_{t}(\omega) \}$, where
$\omega$ labels the possible different ways the jumps might occur.
The statistical operator $\rho_{t}$ associated to such an ensemble
satisfy the following Lindblad-type equation:
\begin{equation} \label{eq:stat}
\frac{d}{dt}\, \rho_{t} \; = \; - \frac{i}{\hbar}\, [H, \rho_{t}]
- \lambda \sum_{n = 1}^{N} \left( \rho_{t} - \int d^{3}x \;
L_{n}({\bf x}) \rho_{t} L_{n}({\bf x}) \right).
\end{equation}

The GRW model and similar models which have appeared in the
literature have been extensively studied (see \cite{cm,pp} for a
review of the subject); in particular, the following three important
properties have been proved:
\begin{itemize}
\item At the microscopic level, quantum systems behave almost
exactly as predicted by standard Quantum Mechanics, the differences
between the predictions of the GRW model and of Quantum Mechanics
being so tiny that they cannot be detected with present-day
technology.

\item At the macroscopic level, wavefunctions of macro-objects
are almost always very well localized in space, so well localized
that their centers of mass behave, for all practical purposes, like
point-particles moving according to Newton's laws.

\item In a measurement-like situation, e.g. of the von Neumann type,
GRW reproduces---as a consequence of the modified dynamics---both
the Born probability rule and the postulate of wave-packet
reduction.

\end{itemize}
Accordingly, models of spontaneous wavefunction collapse provide a
unified description of all physical phenomena, at least at the
non-relativistic level, and a consistent solution to the measurement
problem of Quantum Mechanics.

\subsection*{2.2. Features of the GRW Model}

There are some important properties of the GRW model, which all
non-relativistic collapse models share, and which are relevant for
the subsequent discussion.

\noindent {\bf Non-linearity and stochasticity.} The jump process is
non-linear, since the probability of a jump taking place at {\bf x}
depends on the square norm of the statevector after the hitting. It
is also intrinsically stochastic; the model assumes that Nature is
fundamentally random; needless to say, such a property is important
in order to recover quantum probabilities when measurement
situations are taken into account, but also for other reasons which
will be clear soon.

\noindent {\bf Non-locality.} The model is manifestly non-local; let
us take as an example the following entangled state of two particles
$a$ and $b$:
\begin{equation} \label{eq:epr}
\psi({\bf x}_{a}, {\bf x}_{b}) \; = \; \frac{1}{\sqrt{2}} \, \left[
\psi_{\Delta_{1}}({\bf x}_{a}) \, \psi_{\Delta_{2}}({\bf x}_{b}) \;
+ \; \psi_{\Delta_{3}}({\bf x}_{a}) \, \psi_{\Delta_{4}}({\bf
x}_{b}) \right],
\end{equation}
where $\psi_{\Delta_{1}}({\bf x})$ is a normalized wavefunction well
localized within the region $\Delta_{1}$ of space, and similarly for
the other three terms in (11); $\Delta_{1}$, $\Delta_{2}$,
$\Delta_{3}$ and $\Delta_{4}$ label four regions which are
arbitrarily far away from each other. Let us suppose that an
experimenter $A$ decides to measure the position of particle $a$,
while an experimenter $B$ decides to measure the position of
particle $b$. The full initial state of the two particles plus the
two apparata is:
\begin{equation} \label{eq:epr-app}
\psi_{\makebox{\tiny before}}({\bf x}_{a}, {\bf x}_{b}; {\bf
y}_{\mathcal A}, {\bf y}_{\mathcal B}) \; = \; \frac{1}{\sqrt{2}} \,
\left[ \psi_{\Delta_{1}}({\bf x}_{a}) \, \psi_{\Delta_{2}}({\bf
x}_{b}) \; + \; \psi_{\Delta_{3}}({\bf x}_{a}) \,
\psi_{\Delta_{4}}({\bf x}_{b}) \right]\otimes \phi_{\makebox{\tiny
Ready}}({\bf y}_{\mathcal A}) \phi_{\makebox{\tiny Ready}}({\bf
y}_{\mathcal B}),
\end{equation}
where $\phi({\bf y}_{\mathcal A})$ and $\phi({\bf y}_{\mathcal A})$
denote the (localized) states of, let us say, the pointers of the
two apparata, which initially are both in a state which is ``ready''
for the measurement.

Now, let us suppose that the position of particle $a$ is measured
slightly before that of particle $b$. The dynamics of the GRW model
tells that---because of the spontaneous jumps whose effect is
amplified when a macro-object like a measuring apparatus enters into
play---the final state of the pointer of the apparatus will be
perfectly localized in space and will correspond either to the
outcome ${\mathcal A}: \Delta_{1}$ or to the outcome ${\mathcal A}:
\Delta_{3}$, each occurring with a probability almost identical to
that given by the Born probability rule; let us suppose that the
first possibility occurs. Then, the GRW dynamics implies that the
initial state~\eqref{eq:epr-app}, after the first measurement,
practically reduces to:
\begin{equation} \label{eq:epr-app-aft}
\psi_{\makebox{\tiny after}}({\bf x}_{a}, {\bf x}_{b}; {\bf
y}_{\mathcal A}, {\bf y}_{\mathcal B}) \; = \;
\psi_{\Delta_{1}}({\bf x}_{a}) \, \psi_{\Delta_{2}}({\bf x}_{b})
\otimes \phi_{\makebox{\tiny ${\mathcal A}: \Delta_{1}$}}({\bf
y}_{\mathcal A}) \phi_{\makebox{\tiny Ready}}({\bf y}_{\mathcal B}),
\end{equation}
We see that, because of the (local)  jumps which occurred on the pointer of
the measuring device used by $A$ to measure the position of particle
$a$, also particle $b$ has been almost {\it instantaneously}
localized in space, in this case within $\Delta_{2}$, no matter how
distant $\Delta_{2}$ is from $\Delta_{1}$; in fact, as we see from
state~\eqref{eq:epr-app-aft}, a subsequent measurement of the
position of particle $b$ will give (almost) certainly the outcome
${\mathcal B}: \Delta_{2}$. In a similar way, if the outcome of the
first measurement had been ${\mathcal A}: \Delta_{3}$, then particle
$b$ would have been immediately localized around $\Delta_{4}$, and
this would be confirmed by any subsequent measurement of its
position.

Accordingly, there is a perfect and non-local correlation between
the region where particle $a$ is located after a measurement and the
region where particle $b$ is located by that same measurement done
on its far away partner. Note, however, that the jumps acting on the
pointer (and determining in this way the outcome of a  measurement)
are the consequence of    a perfectly local interaction between the
pointer and the stochastic background which enters in the dynamical
evolution; it is only the entanglement between the two particles
which renders the overall effect fundamentally non-local.

The crucial point to understand is that this non-local feature of
the collapse mechanism is {\it not} a consequence of the fact that
the GRW model is non-relativistic\footnote{Indeed, it would be very
easy to devise a local jump process, even for entangled states,
which in any case would lead to a conflict with quantum mechanical
predictions.}; on the contrary, such a peculiar feature is {\it
necessary} in order to reproduce the {\it quantum correlations} for
EPR states like~\eqref{eq:epr}, which have been confirmed by all
experiments. In other words, after the work of Bell and the
experiments of Aspect which have shown that Nature is fundamentally
non-local, any GRW model (whether relativistic or not) has to embody
such a non-local behavior in order to reproduce quantum
correlations.

\noindent{\bf No faster-than-light.} One might wonder whether the
non-local character of the jump process might be used to send
faster-than-light signals, but in \cite{B} it has been proven that
this is {\it not} possible, and the physical reason is quite simple
to understand: since jumps are intrinsically random, they cannot be
controlled to implement faster-than-light communication, and as soon
as one averages over all possible jumps, their non-local character
vanishes.

Accordingly, like standard Quantum Mechanics, also the GRW model
shares the ``peaceful coexistence'' between relativity and
non-locality, which is one of the lessons we had lo learn from
Bell's inequalities. Indeed, Bell himself has stated \cite{bell}:
``... I am particularly struck by the fact that the [GRW] model is
as Lorentz invariant as it could be in the nonrelativistic version.
It takes away the ground of my fear that any exact formulation of
quantum mechanics must conflict with fundamental Lorentz
invariance''.

The formal aspects  of this nice feature of the theory consists in
the fact that GRW violates Bell's locality condition by violating
outcome independence, just as standard nonrelativistic quantum
mechanics does. Before proceeding it is useful to  recall that
Bell's locality assumption has been proved \cite{PDOD1,PDOD2} to be
equivalent to the conjunction of the two logically independent
conditions, parameter independence and outcome independence. To
clarify the matter let us fix our notation. We will denote by
$\lambda$ all parameters (which may include the quantum mechanical
statevector or even to reduce to it alone) that specify completely
the state of an individual physical system. For simplicity we will
refer to a standard EPR-Bohm setup and we will denote by
$p^{AB}_{\lambda}(x,y;n,m)$ the joint probability of getting the
outcome $x$ in a measurement at $A$ and $y$ in a measurement at $B$.
Obviously we assume that the experimenters at $A$ and $B$ can make a
free-will choice of the directions $n$ and $m$ along which they will
perform their measurements. They can also choose not to perform the
measurement. Bell's locality assumption can be expressed as
\begin{equation}
p^{AB}_{\lambda}(x,y;n,m)=p^{A}_{\lambda}(x;n, *)p^{B}_{\lambda}(y;
*,m),
\end{equation}
where the symbol $*$ appearing on the r.h.s. denotes  that the
corresponding measurement is not performed. As already anticipated,
the above condition has been proved to be equivalent to the
conjunction of the two following conditions:
\begin{equation}
p^{A}_{\lambda}(x;n,m)=p^{A}_{\lambda}(x;n,*);\;\;\;p^{B}_{\lambda}(y;n,m)=p^{B}_{\lambda}(y;*,m)
\end{equation}
and
\begin{equation}
p^{AB}_{\lambda}(x,y;n,m)=p^{A}_{\lambda}(x;n,m)p^{B}_{\lambda}(y;n,m),
\end{equation}
where we have denoted, e.g., by the symbol $p^{A}_{\lambda}(x;n,m)$
the probability of getting, for the given settings $n,m$, the
outcome $x$ at A. The first conditions express {\it Parameter
independence}, i.e., the requirement that the probability of getting
an outcome at $A (B)$ is independent of the setting chosen at $B
(A)$, while the last conditions ({\it Outcome independence})
expresses the requirement that the probability of an outcome at one
wing does not depend on the outcome obtained at the other wing.

We are now in the conditions of discussing briefly this point with reference
to the twined state of total spin 0 of Conway and Kochen:
\begin{equation} \label{eq:sing}
|\phi_{\makebox{\tiny singlet}}\rangle \;  = \; \frac{1}{\sqrt{3}}
\, \left[ |S_{a:n} = +1 \rangle |S_{b:n} = -1 \rangle + |S_{a:n} =
-1 \rangle |S_{b:n} = +1 \rangle - |S_{a:n} = 0 \rangle |S_{a:n} = 0
\rangle \right],
\end{equation}
the states $|S_{a:n} = +1 \rangle$, $|S_{a:n} = 0 \rangle$,
$|S_{a:n} = -1 \rangle$ and the similar ones for particle $b$ being
the eigenstates belonging to the indicated eigenvalues of the spin
component  along an arbitrary direction $n$. Let us now consider
particle $b$. If particle $a$ is not subjected to any measurement,
the probabilities of the two outcomes for  $S^{2}_{b:n}$
are\footnote{As already remarked the GRW model gives practically the
same predictions and has the same effects as standard quantum
mechanics with the wave packet reduction postulate.}:
\begin{equation}
P(S^{2}_{b:n}=1)=\frac{2}{3}, \; \qquad
P(S^{2}_{b:n}=0)=\frac{1}{3}.
\end{equation}
Suppose now that particle $a$ has been subjected to a measurement of
$S^{2}_{a:n}$: the outcome is either 1 or 0;  then
$P(S^{2}_{b:n}=1)=1$ and $P(S^{2}_{b:n}=0)=0$ or viceversa,
according to the outcome which has been obtained. Such probabilities
differ from those of no measurement on particle $a$. This shows that
the theory exhibits Outcome Dependence. However, if one performs a
non-selective measurement on particle $a$, one gets the outcome 1
with probability $\frac{2}{3}$ and the outcome 0 with probability
$\frac{1}{3}$. Since reduction takes place to the states
$\frac{1}{\sqrt{2}}[|S_{a:n} = +1 \rangle |S_{b:n} = -1 \rangle +
|S_{a:n} = -1 \rangle |S_{b:n} = +1 \rangle]$ and $|S_{a:n} = 0
\rangle |S_{b:n} = 0 \rangle$, respectively, the probabilities of
the two outcomes for a measurement on particle $b$ coincide
precisely with those of the case of no measurement. The model then
does not exhibit Parameter Dependence. Concluding: the GRW model
violates locality by violating only outcome independence and, as well
known, since this feature of the theory does not forbid having a
relativistic quantum theory, it therefore does not forbid
relativistic dynamical reduction processes.

\noindent{\bf Stochastic Galilean invariance.}  When dealing with a
theory like the GRW model which contemplates a stochastic evolution,
one has to be careful in specifying the precise meaning of the
theory being invariant under the transformations of a symmetry
group. In the nonrelativistic case this group is naturally
identified with the Galilei group, which, however, in the present
case must be restricted to the so called Galilei semigroup ($GS$)
which contains only forward time translations, since the theory has
a built in arrow of time.

Let us then consider an observer $O$ who  prepares a system in a
state $\vert\Psi^{O}_{0}\rangle$ and lets it evolve, under the
combined effect of the purely Hamiltonian evolution and the random
localization processes up to time $t$, getting the state
$\vert\Psi^{O}_{t}(\omega)\rangle$. Here $\omega$ specifies the
precise localizations processes which have taken place in the
interval $(0,t)$. Obviously, since the localizations are random
processes, the state $\vert\Psi^{O}_{t}(\omega)\rangle$ has a
certain probability $P_{\Psi^{O}_{0}}(\omega)$, which depends on the
initial state and the localizations which took place, of being the
state describing the situation at time $t$.

Let us  consider now another observer $O'$  which is related to $O$
by a transformation $g\in GS$. We know how the states associated by
$O$ and $O'$ are related: $O'$ associates to the system states which
are obtained from those of $O$ by the unitary operator $U_{g}$
implementing the transformation  $g\in GS$ on the Hilbert space of
the system.

We can now make precise the notion of  stochastic invariance. The
theory is invariant for the transformations of the Galilei semigroup
if its dynamics implies that the transformed initial state
$U_{g}\vert\Psi^{O}_{0}(\omega)\rangle$ has precisely the
probability $P_{\Psi^{O}_{0}}(\omega)$ (i.e. the one characterizing
the evolution for $O$) of evolving into the transformed state
$U_{g}\vert\Psi^{O}_{t}(\omega)\rangle$. In ref. \cite{grwrel} this
has been proved to be authomatically true within GRW if the
hamiltonian governing the pure Schr\"odinger evolution is invariant
for the considered transformation, in the usual sense of standard
quantum mechanics. In simple words, the GRW model is stochastically
invariant under the transformations of the Galilei
semigroup\footnote{In a recent nice paper \cite{PO}, the authors
have appropriately stressed the necessity of enriching the pure
formal structure of a theory by what they have denoted as its
Primitive Ontology (PO) and they have mentioned that for the GRW
model one can at least choose two PO's, which they call the Flashes
Ontology and the Mass Density Ontology, respectively (we refer the
interested reader to the paper). Here we remark simply that the
invariance property we have just discussed implies that the GRW
theory can be claimed to be invariant according to the definition
they have used for this purpose: {\it  To say that a theory has a
given symmetry is to say that the possible histories of the PO,
those which are allowed by the theory, when transformed according to
the symmetry, will again be possible histories for the theory, and
the probability distribution on the histories supplied by the
theory, when transformed, will again be a possible probability
distribution for the theory.}}.

\subsection*{2.3. Relativistic GRW Models}

In the past years a great deal of work has been done in order to
formulate a relativistic version of the GRW model, and some
interesting results have been achieved, even though no final, fully
satisfactory model is available yet. To be coincise, it suffices
here to make only a few general remarks, all of which should be
clear from the previous discussion of the GRW model:

\begin{itemize}
\item Any relativistic extension of the GRW model must be
{\bf non-linear} and {\bf stochastic} in order to provide a solution
to the measurement problem of Quantum Mechanics.

\item Such an extension must be also {\bf non-local}, if it aims at
reproducing quantum correlations for EPR types of experiments, in
particular when they are performed at space-like separated regions;
in one way or another, the jump process, even though it is triggered
locally, must ``propagate" practically instantaneously. As a
consequence, the (stochastic) equation for the evolution of the
statevector $|\psi_{t}\rangle$ must be highly non-local.

\item At the same time, the extension must satisfy the {\bf no
faster-than-light} constraint, in order to be compatible with
special relativity. From the previous analysis the strategy one has
to adopt in order to achieve such a result should be clear: due to
stochasticity, only the {\it average} effect of the jump processes,
not the {\it single} realizations of the jumps, can be controlled;
accordingly, the statistical properties of the jump mechanism must
be chosen in such a way that, when the average over all possible
realizations is taken, no non-local feature can be exploited to send
signals at a speed greater than the speed of light. Stated in more
mathematical terms, while the equation for $|\psi_{t}\rangle$ must
be non-local, the equation for the statistical operator $\rho_{t}$
must be local in the sense that no measurement done in any region of
space can in any way alter instantaneously the statistical
distribution of the outcomes with respect to measurements done at
space-like separated regions.

\item Finally, the extension will not be {\bf Lorentz invariant} in
the ordinary sense, but in the {\bf stochastic} sense previously
discussed in connection with the GRW model.
\end{itemize}
Of course, the major difficulty is to put all these ideas into a
consistent model, with a working equation; as already remarked, no
fully satisfactory model has been developed so far, but we want to
draw attention to ref.~\cite{grel}, where a general framework for
relativistic reduction models is presented. In the quoted paper the
analysis is performed with reference to a toy model which does not
assume a specific dynamics for the reduction mechanism (this is
where the difficulty in making the GRW model relativistically
invariant lies, and the reason why this problem is still open), but
it allows one to conclude that there is no reason of principle
forbidding the relativistic reduction program. Within this
framework, following previous ideas of Aharonov and Albert on
relativistic formulations of the postulate of wave-packet reduction
\cite{aa1,aa2,aa3}, the collapse mechanism is supposed to occur {\it
instantaneously} along {\it all} spacelike hypersurfaces crossing
the center of the jump process; in spite of this superluminal
effect, the whole picture is perfectly Lorentz invariant (in the
stochastic sense), it agrees with quantum mechanical predictions, it
does not lead to any contradiction, e.g. it does not allow
faster-than-light signalling and, moreover, different inertial
observers always agree on the outcomes of experiments. Another model
has been recently proposed by R. Tumulka \cite{tum}: it is based on
the multi-time formalism with $N$ Dirac equations, one per particle,
and up to now it works only for non-interacting particles. The open
question is whether it can be consistently generalized to include
also interactions.

\section*{3. THE CONWAY-KOCHEN ARGUMENT AND RELATIVISTIC GRW}

From the previous analysis it should be clear why the argument by
Conway and Kochen does not apply to possible relativistic extensions
of the GRW model: since such an extension must necessarily be
non-local in order to reproduce quantum correlations, it must
violate {\bf FIN}, and the whole argument breaks down. We now repeat
again the argument in the light of a GRW-like description of quantum
phenomena, to see where and how non-locality enters into play, and
also to point out a mistake made by Conway and Kochen in discussing
the collapse process. Let us consider again the singlet
state~\eqref{eq:sing} of total spin 0 for two spin 1 particles, and
let us assume that the two observes $A$ and $B$ perform the two
measurements of the square of the spin in two space-like separated
regions of spacetime; the complete state of the two particles plus
the two apparata is:
\begin{equation} \label{eq:sing-app}
|\psi_{\makebox{\tiny before}}\rangle \;  = \; |\phi_{\makebox{\tiny
singlet}}\rangle \otimes \phi_{\makebox{\tiny Ready}}({\bf
y}_{\mathcal A}) \phi_{\makebox{\tiny Ready}}({\bf y}_{\mathcal B})
\end{equation}

Needless to say, the {\bf FREE} axiom must be true, thus $B$ is free
to measure along any triple of directions. Let us suppose that, with
respect to a given reference frame, he is the first who does the
measurement; he chooses the triple $n,j,k$ and obtains e.g. the
result:
\begin{equation}
S^{2}_{b:n}(j,k;\beta) = 1, \quad S^{2}_{b:j}(n,k;\beta) = 1, \quad
S^{2}_{b:k}(n,j;\beta) = 0.
\end{equation}
Now the crucial point comes. If we require the putative GRW model to
fulfill the {\bf TWIN} axiom, i.e. if we require (and we want to
require it) the model to agree with quantum mechanical predictions,
then we must admit that, as a consequence of the measurement made by
$B$ on particle $b$ - i.e. as a consequence of the jumps acting on
the device used by $B$ in such a way to determine the required
outcome - the {\it whole} state of the two particles must change
almost immediately (and in a non-local manner) from the
state~\eqref{eq:sing-app} to the new state:
\begin{equation} \label{eq:collap}
|\psi_{\makebox{\tiny after}}\rangle \; = \; | S_{a:k} = 0 \rangle |
S_{b:k} = 0 \rangle \otimes\phi_{\makebox{\tiny Ready}}({\bf
y}_{\mathcal A}) \phi_{\makebox{\tiny ${\mathcal B}: S_{b:k} =
0$}}({\bf y}_{\mathcal B});
\end{equation}
only in this way we can be sure that, if $A$ decides to measure the
square of the spin of particle $a$ along the {\it same} three
directions $n,j,k$ chosen by $B$, he will certainly get the same
results which $B$ obtained. We stress it again: a non-local and
almost instantaneous collapse from state~\eqref{eq:sing-app}
to~\eqref{eq:collap} is {\it necessary} in order for the putative
model to account for the quantum correlations involving entangled states.

But now we see that the state of particle $a$ has changed because of
something which happened {\it outside} its backward light-cone (the
jumps acting on the device used by $B$ at a space-like separated
distance), and this change will affect the response of $a$ to spin
measurements. Moreover, the new state of particle $a$ depends on the
choice of directions chosen by $B$ to measure the square of the spin
of particle $b$. Because of this non-local dependence, the argument
of Conway and Kochen does not hold anymore.

In their criticism of relativistic GRW models, Conway and Kochen
mistakenly assume that the jump processes somehow have a local
character, i.e. that only the jumps which occurred in the backward
light-cone can affect the behavior of a particle. But, as we have
seen this is not the right picture: the reduction mechanism, even
though it is triggered by a {\it perfectly local} particle-apparatus
interaction, must have a well-defined non-local character such that
the outcome of the measurement which $A$ performs on particle $a$
{\it does}---indeed, it {\it must}---depend on the outcome of the
measurement performed by $B$ on $b$ (or vice-versa), in particular
on the choice of directions made by $B$, even when the two
measurements occur at space-like separated regions. Accordingly, the
putative model deliberately violates {\bf FIN}, because the jumps
(which Conway an Kochen regard as part of information) propagate
instantaneously. But this does not necessarily generate a conflict
with special relativity since, as previously discussed, such jumps
must be such that they cannot be used to send signals at a speed
greater than the speed of light\footnote{Contrary to Conway and
Kochen, we think that the jump processes should not be regarded as
information, whatever this word precisely means. The reason is that
they cannot be known ahead of time, they cannot be controlled, and
they cannot be used to convey other information ..., i.e., they do
not behave as what is typical of information.}.

Of course, one might wonder how it is possible to reconcile such a
highly non-local behavior of the jump mechanism with special
relativity; as we pointed out in Sec. II, this is still an open
problem, a tentative solution to which has been given in \cite{grel}
and \cite{tum}.

To conclude, because relativistic extensions of the GRW model which
reproduce quantum mechanical correlations must be non-local in one
way or another, the Conway and Kochen argument does not apply to
them: the problem of finding a fully convincing relativistic model
of spontaneous wavefunction collapse remains open. Of course it is a
difficult problem, but there is no reason of principle which makes
it impossible.

\section*{ACKNOWLEDGEMENTS}

We wish to thank S.L. Adler, D. Deckert, D. D\"urr, S. Goldstein, R.
Tumulka and N. Zangh\`i for many useful comments.


\begin{thebibliography}{99}

\bibitem{ck}
J. Conway and S. Kochen, {\it Found. Phys.} (2006), to appear.
ArXiv: quant-ph/0604079v1 (11 Apr 2006).

\bibitem{ks}
S. Kochen and E. Specker, {\it J. Math. Mech.} {\bf 17}, 59 (1967).

\bibitem{hr}
P. Heywood and M.L.G. Redhead, {\it Found. Phys.} {\bf 13}, 481
(1983).

\bibitem{sta}
A. Stairs, {\it Phil. Sci.} {\bf 50}, 578 (1983).

\bibitem{bro}
H.R. Brown and G. Svetlichny, {\it Found. Phys.} {\bf 20}, 1379
(1990).

\bibitem{elb}
A. Elby, {\it Found. Phys.} {\bf 20}, 1389 (1990).

\bibitem{As1}
A. Aspect, P. Grangier and G. Roger, {\it Phys. Rev. Lett.} {\bf
49}, 91 (1982);

\bibitem{As2}
A. Aspect, J. Dalibard and G. Roger, {\it Phys. Rev. Lett.} {\bf
49}, 1804 (1982).

\bibitem{Eb}
P.H. Eberhard, {\it Nuovo Cimento} {\bf 46 B}, 392 (1978).

\bibitem{G}
G.C, Ghirardi, A. Rimini and T. Weber, {\it Lett. Nuovo Cimento}
{\bf 27}, 293 (1980).

\bibitem{Sh}
A. Shimony, {\it Intern. Philos. Quarterly} {\bf 18}, 3 (1978).

\bibitem{B}
J. Butterfield, G. Fleming, G.C. Ghirardi and R. Grassi, {\it Int.
J. Theor. Phys.} {\bf 32}, 2287 (1993).

\bibitem{hem}
D. Hemmick, {\it Hidden Variables and Nonlocality in Quantum
Mechanics} (Ph.D. dissertation, 1996).

\bibitem{grw}
G.C. Ghirardi, A. Rimini and T. Weber, {\it Phys. Rev. D} {\bf 34},
470 (1986).

\bibitem{adl}
S.L. Adler, {\it J. Phys. A}, to appear. ArXhiv: quant-ph/0605072v8
(23 Aug 2006).

\bibitem{cm}
A. Bassi and G.C. Ghirardi, {\it Phys. Rept.} {\bf 379}, 257 (2003).

\bibitem{pp}
P. Pearle, in {\it Open systems and measurement in relativistic
quantum theory}, Lecture Notes in Physics {\bf 526}, H.-P. Breuer
and F. Petruccione eds. (Springer-Verlag, Berlin, 1999).

\bibitem{grwrel}
G.C. Ghirardi, R. Grassi and P. Pearle, {\it Found. Phys.} {\bf 20},
1271 (1990).

\bibitem{bell}
J.S. Bell, {\it Speakable and unspeakable in Quanum Mechanics}
(Cambridge University Press, Cambridge, 1987).

\bibitem{PDOD1}
J. Jarret, {\it Nous} {\bf 18}, 569 (1984).

\bibitem{PDOD2}
P. Suppes and M. Zanotti,
in {\it Logic and Probability in Quantum Mechanics}, P. Suppes ed.
(Reidel, Dordrecht, 1976).

\bibitem{PO}
V. Allori, S. Goldstein, R. Tumulka and N. Zangh\`i, ArXiv
quant-ph/0603027v2 (26 Jun 2006).

\bibitem{grel}
G.C. Ghirardi, {\it Found. Phys.} {\bf 30}, 1337 (2000).

\bibitem{aa1}
Y. Aharonov and D.Z. Albert, {\it Phys. Rev. D} {\bf 24}, 359
(1981).

\bibitem{aa2}
Y. Aharonov and D.Z. Albert, {\it Phys. Rev. D} {\bf 21}, 3316
(1980).

\bibitem{aa3}
Y. Aharonov and D.Z. Albert, {\it Phys. Rev. D} {\bf 29}, 228
(1984).

\bibitem{tum}
R. Tumulka, arXiv quant-ph/0406094v1 (14 Jun 2004).

\end{thebibliography}
\end{document}